\documentclass[12pt,singlecolumn,letterpaper]{article}

\usepackage[utf8]{inputenc}
\usepackage{ifthen}
\usepackage[pdftex]{graphicx} 
\usepackage{url}
\usepackage{latex8}
\usepackage{slashbox}
\usepackage{listings}

\newcommand{\eg}{e.\,g.\ }
\newcommand{\ie}{i.\,e.\ }

\begin{document}

\title{An exploration of CUDA and CBEA for a gravitational wave data-analysis
application (Einstein@Home)}
\author{Jens Breitbart${}^1$ and Gaurav Khanna${}^2$}
\affiliation{${}^1$Research Group Programming Languages / Methodologies\\
Universit\"{a}t Kassel\\
Kassel, Germany\\
\textsf{jbreitbart@uni-kassel.de}\\
${}^2$Physics Department, University of Massachusetts at Dartmouth\\
North Dartmouth, MA, USA \\
\textsf{gkhanna@umassd.edu}
}

\maketitle
\thispagestyle{empty}

\begin{abstract}

\noindent We present a detailed approach for making use of two new computer hardware 
architectures -- CBEA and CUDA -- for accelerating a scientific data-analysis 
application (Einstein@Home). Our results suggest that both the architectures suit 
the application quite well and the achievable performance in the same software 
developmental time-frame, is nearly identical.
\end{abstract}

\section{Introduction}
\label{sec:intro}

\noindent The recent decades have brought a tremendous rise in computer simulations, in
nearly every area of science and engineering. This is partly due to the
development of (Beowulf) cluster computing that involves putting together
``off-the-shelf'' computing units (for example, commodity desktop computers)
into a configuration that would achieve the same level of performance, or even
outperform, traditional supercomputers at a fraction of the cost. Computational
science has benefited and expanded tremendously in the last decade due to rapid
improvements in CPU performance ({\em Moore's Law}) and major price drops due to
mass production and intense competition.

However, a few years ago the computer industry hit a serious {\em frequency
wall}, implying that increasing the processor's clock-rate for gains in
performance could not be done indefinitely, due to rapid increases in power
consumption and heat generation ({\em power wall}). This led all the major
processor manufacturers (such as Intel and AMD) toward multi-core processor
designs. Today, nearly all commodity desktop and laptop processors are
multi-core processors \ie, they combine two or more independent computing cores
on a single die. Thus, manufacturers today continue to pack more power in a
processor, even though their clock-frequencies have not risen (and have
stabilized at around 3 GHz). It is expected that this approach would work well
for the next few years, but both Intel and AMD anticipate that even this
approach will not scale well beyond 8 or 16 cores.

On the other hand, the overall performance of other computing technologies (\eg 
graphics cards, gaming consoles, etc.) has continued to increase at a
rapid rate, thus making general-purpose computing on such devices a tantalizing
possibility. Both Compute Unified Device Architecture~(CUDA)~\cite{cuda} and
Cell Broadband Engine Architecture~(CBEA)~\cite{cell} are new hardware
architectures designed to provide high performance and scaling for multiple
hardware generations. CUDA is NVIDIA's general-purpose software development
system for graphics processing units~(GPUs) and offers the programmability of
their GPUs in the ubiquitous C programming language in conjunction with a set of
libraries for memory management. The Cell Broadband Engine~(CBE), which is the
first incarnation of the CBEA, was designed by a collaboration between Sony,
Toshiba, and IBM (so-called STI). The CBE was originally intended to be
used in game consoles~(namely Sony's Playstation~3~\cite{ps3}) and consumer
electronics devices, but the CBEA itself was not solely designed for this
purpose and has been used in areas such as high-performance computing as well. 

Both CUDA and CBEA are not the only new architecture approaches to overcome a
decline in overall performance gains, but these are expected to be the ones with
the widest distribution. CUDA is available for all NVIDIA GPUs based on the
G80 architecture and its successors, so all currently sold NVIDIA GPUs support
CUDA. Furthermore, Sony has sold over 20 million Playstation~3 game consoles,
each of which has a CBE main processor. Moreover, currently the fastest and the
first petascale supercomputer is the IBM Roadrunner~\cite{roadrunner}, which is
a {\em hybrid} system, built using nearly 13,000 Cell processors and 6,500 AMD
Opteron processors. Based on these successes, an industry standard -- 
OpenCL~\cite{opencl} -- has been proposed by Apple and others, that would allow 
for computing across a variety of different hardware: CPU, GPU, CBE, etc. We 
consider the current distribution of this hardware an important factor for the 
future success of its underlying concepts and programming models, and therefore 
we decided to work with these two architectures.

In this article, we will compare the current state of both CUDA and the CBEA by
modifying the Einstein@Home client application, to enable it to take advantage
of these different hardware architectures. Einstein@Home~\cite{eah} is a
distributed computing project, that uses the computing power volunteered by end
users running its client application, to perform data-analysis tasks for various
gravitational-wave observatories such as LIGO~\cite{ligo} and GEO~\cite{geo}.
The computation performed by this client application can be executed in a data-parallel
fashion, which suits both CUDA and the CBE very well.

This article is organized as follows. Section~\ref{sec:einstein} gives an
overview of the Einstein@Home project in general, and its client application.
The next two sections, introduce the architecture and the software development
system of the Cell Broadband Engine (Section~\ref{sec:cell_arch}) and our
experiences with the development of the Einstein@Home client using it
(Section~\ref{sec:cell_impl}). The following two sections describe CUDA
(Section~\ref{sec:cuda_arch}) and our experiences with GPU architecture
(Section~\ref{sec:cuda_impl}). Finally, in Section~\ref{sec:compare} we compare both
implementations and outline the benefits and drawbacks of CUDA and the CBE and
their respective software development systems. Section~\ref{sec:related} discusses
related work, while Section~\ref{sec:conclusion} provides a summary of this work.


\section{Einstein@Home}
\label{sec:einstein}

\noindent Gravitational wave observatories are currently being built all over the world:
LIGO in the United States, GEO/Virgo in Europe and TAMA in Japan. These will
open a new window onto the Universe by enabling scientists to make astronomical
observations using a new and significantly better medium -- gravitational waves,
as opposed to electromagnetic waves (light). These waves were predicted by
Einstein's theory of Relativity, but have not been directly observed because the
required technology was simply not advanced enough, until very recently
(indirect evidence validating their existence has been available for some
time). 

These observatories generate data at the rate of several tens of GBs per day
and they require highly computationally intensive data-analysis (mainly due to
the fact that the signal-to-noise ratio in the data streams is very low).
Einstein@Home is a BOINC~\cite{boinc} based, public distributed computing
project that offloads the data-analysis associated to these observatories to
volunteers worldwide. The project currently (since 2005) has over 200,000
participants and over 800,000 computers involved, with a strong and sustained
growth pattern. The goal of Einstein@Home is finding gravitational waves emitted
from neutron stars ({\em pulsars}), by running a brute force search for
different waveforms in an extremely large data-set.

We consider the Einstein@Home client a meaningful test application for our
comparision of CUDA and the CBE since from a theoretical viewpoint, its
parallelization is quite straightforward. Furthermore, the application is very
compute intensive, therefore we can expect a high performance gain through
parallelization. The computation of the application can be roughly divided into
two parts -- the so-called {\em F-Statistics} computation, and a {\em Hough-transformation}.
We will only concentrate on the F-Statistics computation in this work, because 
the Hough code is due to be replaced by an alternative, extremely efficient algorithm 
in the near future. We provide below a brief overview of the algorithm and its data 
dependencies; a detailed discussion can be found in~\cite{master}.

Listing~\ref{lst:fstat-code} provides an overview of the F-Statistics code. The
code uses \verb|for each| whenever the loop can be carried out in parallel with
no or minimal changes to the loop body. \verb|+| is used for all commutative
calculations. The parameters of a function denote all the values a function depends
on, however we use ``\textit{...}'' as a reference to all parameters passed to the called
function.  The F-Statistics code consist of multiple nested loops, each looping
through a different part of the input data set, namely: a frequency band, all
used detectors for the observing the gravitational wave and an SFT of each signal.

\begin{lstlisting}[caption={F-Statistics pseudo code}\label{lst:fstat-code}]
T[] ComputeFStatFreqBand(...) {
	T result[];
	int i = 0;
	for each (frequency in frequency_band(...) )
		result[i++]= ComputeFStat(..., frequency);
	return result;
}

T ComputeFStat(...) {
	T result;
	for each (detector in detectors)
		result += ComputeFaFb(..., detector);
	return result;
}

T ComputeFaFb(...) {
	T result;
	for each (SFT in SFTs(frequency))
		result += some_calculations(..., SFT);
	return normalized(result);
}
\end{lstlisting}


In this article we will indicate the current \textit{state} of the calculation, 
as the presently computed upon SFT, detector and frequency -- or more formally speaking: 
the current state is the tuple of $$(frequency, detector, signal)$$ using the 
variable names used in the Listing~\ref{lst:fstat-code}.

For simplicity, Listing~\ref{lst:fstat-code} does not include any data structures 
for storing data. We will provide a brief introduction here (a more detailed 
overview can again be found in~\cite{master}). The data structures used by 
Einstein@Home are based on a deep hierarchy. For example, the data structure 
that is used to store the SFT data for one frequency, is called \verb|multiSFTs|. 
\verb|MultiSFT| has a pointer to a data array, which stores pointers to the SFT 
data for every detector. The SFT data is built of two levels, therefore to finally 
access the data one needs code similar to the one shown in Listing~\ref{lst:DS}. 
The additional two layers within the SFT data itself, are a consequence of how 
the data is stored and processed within the \verb|some_calculations| function. 
We will not go into more details, since the rest consists of calculations that 
must be executed sequentially and are therefore not important for the parallelization.

\begin{lstlisting}[caption={Pseudo code showing the SFT data structure}\label{lst:DS}]
MultiSFT x;
float data = x.data[detector]->data[SFT]->data[i]->data[j];
\end{lstlisting}


\section{Test cases}

\noindent The various measurements presented in this article were done with two different kinds 
of data-sets that not only differ in the overall runtime, but also in their memory 
requirements and performance characteristics. The \textit{small test case} is based on 
a data-set used to check if the Einstein@Home client application itself is producing correct 
results. It uses a small number of frequencies in contrast to a full Einstein@Home 
work unit. In the small data set case, the F-Statistics takes nearly 90\% of the overall 
runtime, while in the {\em full work unit case} its share is nearly 50\%. A full work unit 
consists of multiple, so-called \textit{sky points}. The calculation done for one sky point 
consists of both the F-Statistics and the Hough transformation and the runtime 
required to compute one sky point is nearly identical to the one required 
for the other sky points. For this work, we therefore do not measure the runtime of a 
full work unit, but only measure the time required to calculate one sky point.


\section{Cell Broadband Engine}
\label{sec:cell_arch}

\noindent The CBE is a totally redesigned processor, that was developed collaboratively by
Sony, IBM and Toshiba mainly for multimedia applications. This processor has a
general purpose (PowerPC) CPU, called the PPE (that can run 2 software threads
simultaneously) and 8 special-purpose compute engines, called SPEs available for
raw numerical computation. Each SPE can perform vector operations, which implies
that it can compute on multiple data, in a single instruction (SIMD). All these
compute elements are connected to each other through a high-speed interconnect bus 
(EIB). Note that the design of this processor is very different from traditional 
multi-core processors. In a certain sense, the CBE's design is somewhere between 
a general-purpose CPU and a specialized GPU (as described in Section~\ref{sec:cuda_arch}). 
It can therefore be considered as a {\em hybrid} technology, having the advantages of 
both these architectures. The outcome of this distinctive design is that a single, 
3.2 GHz (original - 2006/2007) CBE has a peak performance of over 200 GFLOPS in
single-precision floating point computation and 15 GFLOPS in double-precision.
It should be noted that the current (2008) release of the CBE, called the
PowerXCell, has design improvements that bring the double-precision performance
up to 100 GFLOPS. 

We will not attempt to go into much more detail concerning the CBE's design
here, rather we will simply point out one unique feature that addresses the
issue of the {\em memory wall} that is common to all current computer hardware.
The {\em memory wall} refers to the large (and increasing) gap between processor
and memory performance, causing the slower memory speeds to become a significant
bottleneck. The current state-of-the-art approach to combat this
issue has been to include large cache sizes (several MBs) on the processor chip.
However, this takes away valuable space (for compute elements) on the processor
die, and thus can result in only a marginal, overall performance increase. A key
feature of the the CBE is its unique ability to {\em interleave} computation and
data access. Therefore, it is possible for the programmer to overlap memory
access and the actual computation (``double buffering''), in order to hide the
time it takes to access memory. It is this mechanism that allows the CBE to
break through the {\em memory wall} and perform very efficiently, even for
computations that have a large memory footprint. It is also partly for this
reason, that the CBE can reach a ``real world'' application performance that is
nearly 100\% of its theoretical peak performance~\cite{realperformance}. The
parallel programming model on CBEA allows for the use of SPEs for performing
different tasks in a workflow (``task parallel'' model) or performing the same
task on different data (``data parallel'' model). We use the data parallel model 
in our implementations.

One (software) challenge introduced by this new design, is that the programmer
has to explicitly manage the memory transfer between the PPE and the SPEs. The
PPE and SPEs are equipped with a DMA engine -- a mechanism that enables data
transfer to and from main memory and each other. Now, the PPE can access main
memory directly, but the SPEs can only directly access their own, rather limited
(256KB) local store. This poses a challenge for some applications,
including the Einstein@Home client application that we are considering in this
article. However, compilers (IBM XLC/C++) are now available that enable a
{\em software caching} mechanism that allow for the use of the SPE local store as a
conventional cache, thus negating the need of transferring data manually from
main memory. Another important mechanism that allows communication between the 
the different elements (PPE, SPEs) of the CBE is the use of mailboxes. These 
are special purpose registers that can be used for uni-directional communication.  
Each SPE has three mailboxes -- two outbound, that can hold only a single entry, 
and one inbound, that can hold four entries. These are typically used for 
synchronizing the computation across the SPEs and the PPE, and that is primarily 
how we made use of these registers as well. Details on our specific use of these various 
aspects of the CBE for the Einstein@Home client application appear next section 
of this article.

We will end this section with a brief remark on the future roadmap of the CBE as
released by IBM~\cite{roadmap}. STI is on track to deliver the next generation
Cell processor, with 2 PPE elements and 32 SPEs, in 2010. This processor alone
is expected to be able to deliver a peak performance in the teraflops range. This
roadmap is very strong and the CBEA clearly has longevity, including strong
software support from major corporations like IBM and Sony.

\section{Implementation on the Cell Broadband Engine}
\label{sec:cell_impl}

\noindent As it can be seen in Listing~\ref{lst:fstat-code} the F-Statistics calculation
consist of multiple nested loops that can be carried out in parallel. All loop
iterations in the function \verb|ComputeFStatFreqBand| can be carried out with no
changes, whereas both the loops in the functions \verb|ComputeFStat| and
\verb|ComputeFaFb| execute a reduction. We parallelized the F-Statistics code by
parallelizing the loop in \verb|ComputeFStatFreqBand|. We did this because parallelizing the
outer loop limits the overhead generated by the parallelization and the number
of frequencies is an order-of-magnitude higher than the number of detectors or the 
SFTs. The parallelization is done by equally assigning a number of frequencies to
the SPEs independently, and having each SPEs calculate the results for the assigned
frequencies. The PPE is only used for synchronizing the SPEs -- the PPE tells
the SPEs when to start with the calculations and waits for all the SPEs to finish
their calculations before continuing with the Einstein@Home application. The code
executed by the SPE itself is a copy-and-paste of the original code, except for
the modification that we describe below.

We developed two F-Statistics implementations for the CBE. In the first
implementation, we manually manage transfers to and from the local store of the
SPEs, whereas the second implementation relies on the software cache
implementation provided by IBMs XLC/C++ compiler. In this section, we will first
describe the two implementations and discuss the benefits of each later. We will refer 
to the first implementation as \textit{DMA-Fstat}, whereas the second 
implementation will be called \textit{EA-Fstat}.

As we stated before, DMA-Fstat uses manual DMA data transfers, therefore the first step of the
development for this implementation was to make sure that the data, that must be
transferred to the local store, complies with the memory alignment requirements
for the DMA transfers of the CBE. The IBM Cell SDK includes special \verb|malloc|-like
functions, that return memory which is properly aligned for DMA transfers. 
Imposing the memory alignment requirements for all 
data structures was done by modifying Einstein@Home client \verb|malloc| function 
wrappers, so they call the special \verb|malloc| functions provided by the SDK. 
Measurements show that using the SDK's malloc function for all memory allocations, 
does not cause any performance issues, so we decided to allocate all data 
structures with DMA-complying memory alignments, even though it is required only 
for data structures that need to be transferred to the SPE local store. Furthermore, 
identifying the part of the application, where the data structures used by the SPEs 
are allocated would require a more detailed knowledge of the Einstein@Home client 
than we currently have.

The DMA-Fstat build of the client is based on the well known thread programming paradigm. 
The PPE creates multiple threads, each of which is used to control a single
SPE. After the threads are created, the PPE inputs the data structure addresses
used by F-Statistics into the mailboxes of the SPEs. This communication is also
used to notify the SPEs to begin work. After the SPEs have received
the addresses, they use DMA transfers to get all data required for the complete computation. 
We cannot use double buffering because the data that is needed for the calculation is 
computed on the fly for most data structures. We could have implemented double buffering 
for some data structures, but we did not do so, because DMA-Fstat cannot be used 
for a full work unit anyway (explained below). Since we did not use double buffering, 
that diminishes the possible performance gain we could achieve with this implementation. 
Moreover, the need to transfer all data at the beginning of the calculation, in conjunction 
with the rather small size of the local store available on the SPEs limits the amount of 
data that can be processed by each SPE. DMA-Fstat works well for the small data set case, 
but is unable to process the full data set. After the data is processed, the SPEs write 
their results back to main memory by using DMA transfers and place a ``work finished'' 
message in the mailbox. The PPE waits until all SPEs have placed this message in their
mailbox, before the Einstein@Home client is executed any further.

We developed our second F-Statistics implementation~(EA-Fstat) to no longer be
limited by the amount of data that can be processed. EA-Fstat relies on the SPE
software cache implementation of the XLC/C++ compiler, that freed us from
manually transferring data to the local store. Furthermore, we did not need
to guarantee any memory alignment for the data that must be accessed by the SPE.
We only needed to guarantee that in the SPE code all pointers, pointing to main
memory are qualified with the \verb|__ea| qualifier. Since the data structures
used by the Einstein@Home client are deeply pointer based, this modification
took some time, but was far less intrusive than changing the memory alignment of
all data structures, as was done in DMA-Fstat. The initial communication
is done identically as in DMA-Fstat, meaning that the addresses of the data
structures in main memory are sent to the SPE mailboxes. These addresses are
assigned to \verb|__ea| qualified pointers and then used as if they point
to locations in the SPE's local store. The synchronization of SPE and PPE is 
again done identically to that of DMA-Fstat, however before the SPE sends out 
the ``work finished'' message in the mailbox, the SPE writes back the data stored 
in the local store cache. The cache write back is done by manually calling a 
special function.

The benefit of EA-Fstat compared to DMA-Fstat is that the developer no longer
has to worry about the size of the local store and can automatically benefit from
larger local store in future hardware generations. Furthermore, relying on the
software cache reduces the complexity of the SPE program code: DMA-Fstat
required 122 lines of code (not counting comments) consisting of memory
allocation and pointer arithmetic, whereas the EA-Fstat implementation only
consists of 9 library function calls, that read data structure addresses out
of the mailboxes.

\begin{table}[tbp]
\centering
\caption{The time required for the CBE client running the small test case.}
\begin{tabular}{l l l}
\hline
PPE & DMA-Fstat & EA-Fstat \\ 
\hline \hline
8:53 min & 2:30 min & 2:34 min \\
\hline
\end{tabular}
\label{tab:cell-small}
\end{table}

Table~\ref{tab:cell-small} shows the runtime of the original Einstein@Home client using 
only the PPE in comparision with both our new implementations, for the small work unit case. 
These measurements were performed by running the codes on a Sony Playstation 3, that 
allows us to use a maximum of 6 SPEs. Both DMA-Fstat and EA-Fstat use only 3 SPEs, 
since the small amount of work of the small test case does not scale well with additional 
SPEs. We can see that the PPE is clearly outperformed by the clients that use the SPEs. 
This should come at no surprise since those clients run the F-Statistics calculation in 
parallel on multiple SPEs without any need for synchronization. Furthermore, we can store all 
the data that is required for the F-Statistics computation in the fast local store of the SPE, 
thus there is no need to access slow main memory. The comparision of both DMA-Fstat and EA-Fstat 
shows the performance hit of using the software cache mechanism is only about 2.5\% in our 
small test case. The low performance loss is partly due to the fact that we did not use double 
buffering, which would have likely increased the performance of DMA-Fstat.

We show the performance of our EA-Fstat solution in Table~\ref{tab:cell-full}. When running 
a full work unit the EA-Fstat client cannot outperform the PPE version as well as it did 
in the small data-set test case. When running the client with 6 SPEs, the client finishes 
in about 59\% of the runtime of the PPE-only client -- {\em we gain a factor of 1.7 in overall 
application performance, by making use of the CBE architecture}. The reduced performance gain 
does not seem to be a result of the size of the used software cache -- halving the cache size 
to 64KB reduces the performance by 2\%. The performance of the EA-Fstat client is limited by 
the runtime required by the Hough transformation, which turns out to be the performance 
bottleneck when running the client with 6 SPEs. When considering the F-Statistics computation 
alone, {\em the performance improved by a factor of about 5.5 upon using 6 SPEs} -- F-Statistics 
requires less than two minutes of the overall runtime.

\begin{table}[tbp]
\centering
\caption{Time per sky point for the CBE client for a full work unit case.}
\begin{tabular}{lllllll}
\hline
    &  \multicolumn{6}{c} {Processing elements} \\
\hline \hline
PPE & \multicolumn{6}{c}{SPEs} \\
\hline
 & 1 & 2 & 3 & 4 & 5 & 6 \\
\hline
22 min & 20 min & 16 min & 14.5 min & 13.75 min & 13.5 min & 13 min\\
\hline
\end{tabular}
\label{tab:cell-full}
\end{table}

The best overall performance of the Einstein@Home client on the CBE platform is probably 
achieved by running multiple clients on one system, so the PPE's ability of running 2 software 
threads can be used and all SPEs are kept busy, even when one client is currently executing the 
Hough transformation. Our experimentation suggests that one can gain an additional 30\% in overall 
application performance in this manner. Recall that the Hough code is soon due to be replaced by 
an extremely efficient algorithm -- when that happens, we expect to gain over a factor of 5 in 
the overall application performance. 

\section{CUDA architecture}
\label{sec:cuda_arch}

\noindent CUDA is a general-purpose programming system only available for NVIDIA
GPUs and was first publicly released in late 2007. CUDA requires a GPU
that is based on NVIDIA's so-called G80 architecture or one of its successors.
Through CUDA, the GPU~(called \textit{device}) is exposed to the CPU~(called
\textit{host}) as a co-processor with its own memory. The device executes a
function~(called \textit{kernel}) in the SPMD model, which means that a
user-configured number of threads runs the same program on different data.
Threads executing a kernel must be organized within so called \textit{thread
blocks}, which may consist of up to 512 threads; multiple thread blocks are
organized in a \textit{grid}, which may consist of up to $2^{17}$ thread blocks.
Thread blocks, however, are not there purely for organizational purposes, they also play
an important role related to performance. Each thread block is always
scheduled onto one so-called \textit{multiprocessor} of the device. A single 
multiprocessor consists of 8 processors. The number of multiprocessors of a
device depends on hardware used with the current maximum of multiprocessors on a
single device is 30. If a kernel uses only one thread block, only one
multiprocessor of the device is used and most of the devices' processing power
is not utilized. Multiple thread blocks can be assigned to the same multiprocessor, 
thereby, they equally share the resources of the multiprocessor. Thread blocks are 
important for algorithm design, since only threads within a thread block can be 
synchronized. NVIDIA suggests having at least 64 threads in one thread block and 
up to multiple thousands of thread blocks -- more threads than the device has
processors -- to achieve high performance. Threads within thread
blocks can be addressed with one-, two- or three-dimensional indexes; thread
blocks within a grid can addressed with one- or two-dimensional indexes. We call the 
thread index \verb|threadIdx| and the dimensions of the thread block x, y and z (therefore 
\verb|threadIdx.x| is the first dimension of the thread index). We refer to the thread block 
index as \verb|blockIdx| and use the same names for the dimensions.

From the host's point of view, kernel invocations are asynchronous function
calls. Synchronisation between host and device are done explicitly by calling a
synchronization function, or implicitly when the host tries to access memory on
the device. In both cases, synchronization takes the form of a barrier that
blocks the calling host thread until all previously called kernels are finished.

In contrast to main memory used by the CPU, its GPU counterpart -- called
\textit{global memory} -- is not cached and accessing it costs an order-of-magnitude 
more than most calculations. For example, 32 threads require 400 - 600
clock cycles for a read from global memory, whereas an addition executed by the
same number of threads takes only 4 clock cycles. The high cost of accessing
global memory is another incarnation of the so-called {\em memory wall} that also was
an important factor for the design of the CBE (see Section~\ref{sec:cell_arch}).
However, NVIDIA chose another way to circumvent this problem. The device uses
an efficient thread scheduler that uses the massive parallelism approach of the
device to hide the latency by removing threads that just issued a global memory
read from its processor and scheduling a thread that is not waiting for data.
This is one of the reasons why the device requires more threads than there are
processors available to achieve good performance. Furthermore, CUDA provides a way
of directly reducing global memory accesses, by using a special kind of memory
called \textit{shared memory}. Shared memory is fast memory located on the
multiprocessors of the device itself and is shared by all threads of a thread
block. Accessing shared memory costs about 4 clock cycles for 32 threads and may
be used as a cache, which must be managed by the developer. However, using
global memory cannot be avoided, because it is the only kind of memory,
which can be accessed by both the host and the device. Data that is stored in
main memory must be copied from main memory to global memory by a CUDA memcopy
function call, if it is needed by the device. Results of a kernel that need to be
used by the CPU must be stored in global memory and the CPU must issue a memcopy
from global memory to main memory to use it. All transfers done by CUDA memcopy
functions are DMA transfers and have a rather high cost of initialization and a
rather low cost for transferring the data itself.

\section{Implementation using CUDA}
\label{sec:cuda_impl}

\noindent The development of the CUDA based F-Statistics was an evolutionary process. We
developed three different versions, each solving the problems that emerged in
the previous version. 

We decided to start our work by porting the innermost function of the
F-Statistics (\verb|ComputeFaFb|) to the device. Porting the code to be executed
by the device did not require any difficult work. Our implementation uses one thread
per SFT and only one thread block. The number of SFTs in our data sets are less
than the maximum number of threads in one thread block. Since we use one thread
per SFT, we calculate all the SFTs of one detector (and therefore, all SFTs processed
by one call of \verb|ComputeFaFb|) in parallel. More formally speaking, we
calculate the states $(frequency, detector, threadIdx.x)$ in one kernel call,
with \verb|threadIdx.x| being the x-dimension of the block local CUDA thread
index.

The reduction, which is part of \verb|ComputeFaFb|, is implemented by using
synchronized access to shared memory, which is possible since all threads are in
one thread block. Reductions, that include multiple thread blocks are difficult
to achieve with this hardware, since one cannot synchronize threads of different
thread blocks or support hardware floating point atomic operations. Therefore,
we did not implement a parallel reduction.  A parallel reduction example is 
provided by NVIDIA~\cite{CUDA:reduct}, so we can easily replace our implementation 
in the future. In our implementation, the first thread collects all the data from 
the other threads, calculates the result and writes it to global memory. The 
performance of this first version is not limited by the performance of the 
calculations done at the device, but by the device memory management done by the 
host. We refer to this version as \textit{version~1}.

The concept behind version~1 was chosen to start our work in an easy way, and 
not necessarily provide better performance than the CPU version. Nonetheless, we
expected the overall performance to be half that of the host-based
implementation, but version~1 takes more than 70 times the runtime of the CPU
version! Measurements show that almost 95\% of the runtime is spent on 
device memory management and about 70\% on copying the data to and from
the device. This problem arises because version~1 uses the same data
structure as the original CPU version, which consists of a large number of small
memory blocks (the exact number depends on the data). To transfer the data
structure directly to the device, version 1 needs to issue an DMA transfer to
global memory for every memory block. Since a DMA transfer has a rather high
cost of initialization and a rather low cost for transferring a data element
itself, this data structure design cannot provide high performance when used in
conjunction with CUDA.

Our second implementation~(\textit{version~2}) still calculates the states $(frequency,
detector, threadIdx.x)$ in one function call, but uses a newly developed data
structure. The new data structure was designed to directly benefit from the way
DMA memory transfers are performed. In the kernel, the 
data is typically read once from a known position that only depends on the current
state of the calculation. The requirements of both fast memory transfer and fast
access to an element at an arbitrary but known position are satisfied by simply
using an array.  An array is guaranteed to be stored in one contiguous memory
block and each element can be directly accessed by using an index, its address
or a combination of both \ie an offset based on an address pointing to one
element in the array. Thus, our new data structure is an aggregation of a small
number of arrays.

These arrays can be grouped in two types. One type stores data used to
calculate the result of the F-Statistics and is therefore called \textit{data
arrays}. The other type stores pointers pointing to elements inside the data
arrays -- we refer to this type as \textit{offset arrays}, since the pointers
stored inside these arrays are used as an offset to access data the way we have
just described. The creation of this data structure itself is rather simple. The
original small memory regions are all padded into a data array in an ordered
fashion. The elements in the offset arrays are used to identify the old memory
regions. The new data structure is only used by the device, the code executed at
the host continues to be the previous one.

Performance measurements of version~2 show that our new data structure solves
the problems of version~1. Memory management is still a measurable amount but is
no longer the performance bottleneck. Version~2 has approximately the same performance
as the original CPU version. Next, we decided to change our kernel call concept, as 
used in version~1, to utilize multiple multiprocessors of the device and thus
improve performance (we had repeatedly issued kernel calls
with only one thread block so far). We next describe how we replaced these multiple
calls with just one kernel call using multiple thread blocks.

Let us first revise the important details of Listing~\ref{lst:fstat-code},
before we explain how we implemented the final version of F-Statistics with CUDA. 
If we want to calculate the F-Statistics for a frequency band, we have to port 
the functions \verb|ComputeFStatFreqBand| and \verb|ComputeFStat| to the device.
\verb|ComputeFStat| contains a reduction of the values calculates by
\verb|ComputeFaFb| -- the function we ported to the device in step one. The
results of \verb|ComputeFStat| are written to different memory addresses.

As previously described, we use a fairly simple way to calculate the reduction
done in \verb|ComputeFaFb|, which is possible only because all threads are in
the same thread block. Our analysis of the used data sets showed that we could
continue with this approach. The number of threads required to calculate
\verb|ComputeFaFb| for all detectors is still small enough to have one thread
block calculate them. For easy readability of our code we use two dimensional
thread blocks. The x-dimension still represents the calculated SFT, whereas the
y-dimension is now used to represent the detectors. Version 1 calculates the
states $(frequency, detector, threadIdx.x)$ with one kernel call, whereas our
re-worked approach calculates $(frequency, threadIdx.y, threadIdx.x)$ with one
kernel call. Even though this approach uses more threads, it still uses only one
thread block per kernel call.

The next step is to use multiple thread blocks to calculate different
frequencies \ie to calculate the states $(blockIdx.x, threadIdx.y,
threadIdx.x)$ with one kernel call. Conceptually this can be achieved with
relative ease, since the calculations of one frequency are independent from one
another. Nonetheless, this concept still results in some overhead work that may
not be very obvious at first. For example, the number of signals is not constant
across all frequencies, but the dimension of all thread blocks is identical. To
solve this problem, we first scan the data-set for the maximum number of signals
found in a frequency of a given frequency band and thereby determine the
x-dimension of all thread blocks. This approach results in idle threads for all
the other thread blocks, but our evaluation of the dataset shows only a small
fraction of idle threads -- typically, 6 threads idle per x-dimension, with 
the maximum being 10. This is our final implementation of the Einstein@Home
CUDA application, which we will refer to as \textit{version~3}. The development 
of all three versions was done on a device, that did not support double precision 
floating point operations, which is required by Einstein@Home to produce correct 
results. This client therefore, does not use double precision and the results 
are of no scientific value.

\begin{table}[tbp]
\centering
\caption{Time measurements of the CUDA based client.}
\begin{tabular}{ lll }
\hline
test case& CPU & GPU \\ 
\hline \hline
small & 2:19 min & 1:22 min\\
\hline
full (per sky point) & 11:00 min & 7:00 min\\
\hline
\end{tabular}
\label{tab:cuda_perf}
\end{table}

Upon running our implementation on a test system with a GeForce GTX 280 and two AMD Opteron 
270 (2 GHz) {\em the CUDA version performs about 1.7 times as fast as the CPU version for the 
small test case. When run on a full work unit, the CUDA client performs about 1.6 times as 
fast as the CPU version}. Note that both the CPU and the GPU version only use one core when 
running the Einstein@Home client. {\em Considering only F-Statistics, the performance was improved 
by more than a factor of about 3.5} -- the F-Statistics calculation only takes about 1.5 minutes.

\section{Comparision of CUDA and the CBE}
\label{sec:compare}

\noindent One of the most important factors when developing for the CBEA is the size of local store, 
especially when the developer needs to manually manage the data storage in the local store. 
The data required by our application does not fit in the local store. An easy way to overcome 
this is to use a software cache -- however, by using this technique, the developer loses the 
chance to utilize double buffering, which is one of the most important benefits of the CBEA. 
Using CUDA required us to change the underlying data structure of our application, which we 
could also use with the CBEA. In our case the data structure redesign was rather easy, however 
in other cases this may be more problematic. The main benefit of using CUDA is the hardware, 
that increases performance at a much higher rate than multi-core CPUs or the CBEA. Furthermore, 
software written for CUDA can easily scale with new hardware -- at least when a high number of 
thread blocks is used. However, writing software that achieves \textit{close-to-the-maximum} 
performance with CUDA is very different when compared to most other programming systems, since 
thread synchronization or calculations are typically not the time consuming parts, but memory 
accesses to global memory are. We did not optimize any of our implementations to the maximum 
possible performance, instead invested a similar amount of development time into both the CBE 
and the CUDA client. Therefore, our performance measurements should not be considered as what 
is possible with the hardware, but rather what performance can be achieved within a reasonable 
development time-frame.

\section{Related work}
\label{sec:related}

\noindent Scherl et al.~\cite{CT-Reconstruction:CUDA} compare CUDA and the CBEA for the so called 
FDK method for which CUDA seems to be the better option. Christen et al.~\cite{multicore-stencil} 
explore the usage of CUDA and the CBEA for stencil-based computation, which however does not give 
a clear winner. In contrast, our work does not strive for the maximum possible performance and relies 
on very low cost hardware.

\section{Conclusion / Future work}
\label{sec:conclusion}

\noindent The main outcome of our work presented in this article, is that new upcoming
architectures such as CBEA and CUDA have very strong potential for significant
performance gains in scientific computing. In our work, we focussed on a specific,
data-analysis application from the gravitational physics community, called Einstein@Home.
This is amongst the largest BOINC-based, public distributed computing projects.
Using these architectures, we successfully accelerated by several fold, one of the
two computationally intensive routines of the Einstein@Home client application. Our
final CBEA and CUDA implementations yield comparable performance, although the development
process, software tools, challenges faced, etc. relevant to these two architectures were
completely different. In this article, we described our experiences with this
software development in great detail.

From the point of view of the Einstein@Home project, we expect our work to have
a major impact. The CUDA based GPU client will ultimately enable the project to
harness the vast computing power currently inaccessible in consumer desktop and
laptop computer graphics cards. When the GPU client is deployed, we anticipate the
total computing power of the project to increase substantially. The CBE client
would similarly be extremely helpful to the project. There are currently 20 million
Playstation 3 gaming consoles sporting the Cell processor that could potentially
compute for Einstein@Home once our client application is deployed by the project.
That would likely increase the overall computing power of Einstein@Home many fold. It
is worth noting that the release of a similar Playstation 3 client enabled
Stanford's Folding@Home to quantum leap into the petascale regime in 2007~\cite{folding}.

In the future, we expect to complete the final development of the CBEA and GPU
clients and address some of the issues they currently have, as we have described in
this article. Then we will work closely with the Einstein@Home project team to integrate 
our CBEA and GPU clients with the various currently operational CPU clients, for a 
full-scale deployment.

\section*{Acknowledgment}

\noindent The authors would like to thank NVIDIA and Sony for providing the hardware that was 
used for development, testing and benchmarking our codes. GK would also like to acknowledge 
support from the National Science Foundation (NSF grant number: PHY-0831631).

\end{document}